\documentclass[12pt,thmsa]{article}
\usepackage[]{sw20jart}

\input{tcilatex}
\input tcilatex
\begin{document}

\title{Is the EPR paradox a paradox?}
\author{A. Tartaglia, Dip. Fisica, Politecnico, Corso Duca degli Abruzzi 24, I-10129
Torino, Italy; e-mail: tartaglia@polito.it}
\maketitle

\begin{abstract}
The EPR paradox and the meaning of the Bell inequality are discussed. It is
shown that considering the quantum objects as carrying with them
''instruction kits'' telling them what to do when meeting a measurement
apparatus any paradox disappears. In this view the quantum state is
characterized by the prescribed behaviour rather than by the specific value
a parameter assumes as a result of an interaction.
\end{abstract}

\section{Introduction}

A famous paper by Einstein, Podolsky and Rosen \cite{epr} appeared in 1935
in the Physical Review gave rise to a live debate on the meaning and the
foundation of quantum mechanics, lasting up to our days. Einstein, Podolsky
and Rosen (EPR) pointed out that something must be missing in the quantum
theory because applying it to a thought experiment they analysed, led to a
paradox or at least to a contradiction with the Heisenberg uncertainty
principle. Three decades later, J. Bell proved his celebrated inequality 
\cite{bell} and applying it to the EPR experiment concluded that in some
cases quantum mechanics violates it and that its violation would imply a
break down either of the principle of reality or of the principle of
locality (provided one would not be inclined to give up causality).
Different variants of the EPR experiment have been considered such as the
one proposed by Bohm \cite{bohm} using pairs of spin $1/2$ particles
prepared in a singlet state. Finally A. Aspect and his group, in the 80's,
actually performed a series of EPR type experiments measuring the
polarization of pairs of photons produced in an SPS atomic cascade \cite
{aspect}. They verified that the result of the experiments was correctly
predicted by quantum theory and that in some configurations Bell's
inequality was actually violated.

All this strengthened the idea that quantum mechanics is indeed paradoxical
in character and revived a debate whose ingredients are mainly the supposed
nonlocality of quantum phenomena, if not the existence of some sort of
superluminal interaction between distant quantum objects.

Now what is rather surprising in this story is that the so called EPR
paradox appears as being mainly due to a misuse of the Heisenberg principle
which should have been seen since the very beginning. A vague idea of this
may be found in the answer by Bohr \cite{bohr} to EPR's paper. 

The claim that the paradox was not there and simply was a consequence of an
incorrect application of the rules of probability was first issued by L.
Accardi \cite{accardi1} who developed an axiomatic approach to what he calls
quantum probability \cite{accardi2}.

In the present paper I shall show why the EPR paradox is not a paradox and
why the result of Aspect's experiments does not imply any violation of the
principle of locality. In the next section the basic properties of a quantum
object or system are reviewed. In the third section the EPR type experiments
are analysed, then section 4 contains a discussion of the meaning of the
Bell inequality; finally section 5 summarizes the whole paper.

\section{Quantum objects}

A quantum object is something that can exist in a set of different physical
states. Mathematically these states can be eigenstates of a number of
different operators. Practically the effect of any quantum operator
corresponds to the interaction with a given apparatus which induces the
object to enter one of the eigenstates of the corresponding operator.

We know that operators exist which are mutually incompatible in a sense that
they do not admit common eigenstates. The commutator of two such operators
differs from zero: if $\widehat{A}$ and $\widehat{B}$ are the two operators,
applying them in the order $\widehat{A}$ $\widehat{B}$ leaves the system in
a state different from the one in which it is left applying them in the
order $\widehat{B}\widehat{A}$.

All this means that when the system is in an eigenstate of $\widehat{A}$,
the physical quantity associated to that operator assumes one of its
eigenvalues, say \textit{a, }but the quantity associated with $\widehat{B}$
is undefined. Conversely, when the system is in an eigenstate of $\widehat{B}%
,$ the quantity it supports assumes the eigenvalue \textit{b}, but now the
quantity associated with $\widehat{A}$ is undefined.

This is of course the Heisenberg principle. It is appropriate to stress that
there is nothing mysterious, exotic or peculiar to quantum physics in this.
It has simply been stated, in a refined and mathematically precise way, the
fact that anybody may be standing or seated, but nobody may simultaneously
be standing and seated.

\section{The EPR paradox}

In their thought experiment EPR consider a couple of  particles moving in
opposite directions and so prepared that their total momentum is zero. If we
measure the momentum of one of them we know the momentum of the other 
\textit{without measuring it}. Now if we determine the position of the
second particle, we end up knowing both position \textit{and} momentum of
one and the same object, thus contradicting the Heisenberg principle.

Actually performed experiments use rather pairs of objects prepared in a
singlet state or photons emitted in an SPS atomic transition, so that their
spins are antiparallel (first case) or polarization vectors equal (second
case). Once the two objects are well separated one from the other each of
them is tested, using an appropriate analyser, for the spin component or
polarization state along a given direction. Now using the EPR\ argument, if
the measurements are performed along different directions (for instance
orthogonal directions) we conclude that the spin components or polarization
states of one object along different directions can be simultaneously
determined, in one case by direct measurement and in the other by a
counterfactual but (apparently) sound inference.

The week point in this is the idea of ''element of reality'' (to use EPR's
words) whose value is being tested or inferred. Einstein (and many others
after him) think of the experimentally testable properties of a quantum
system as being things carried around by the system: they are objectively
there, measure them or not. Now this idea appears to be in contrast with
experience, because of the ''paradoxes'' it generates.

Actually the situation is different. We may think that a quantum object
(which is well defined and existent, independently from the fact of being
under measurement or not) carries its properties as an ''instructions kit''
telling it what to do when encountering a measuring apparatus or in general
undergoing an interaction with something else.

In Aspect's experiments the two photons, at the act of separating, receive,
each of them, their ''instructions'' (whose content depends on the way they
are prepared). When they meet an analyser they behave accordingly: the
realism is save and Einstein's locality is save too. If the two objects meet
two experimental sets of the same kind so are the results (in case of spin
components, they may have the same or opposite sign, depending on the
initial preparation). If however one of the members of the pair is led to
interact with a different type of apparatus, its ''instruction card'' tells
it to behave differently.

Now the counterfactual inference that, in the EPR argument, lead to the
''paradox'' tells us what the object under attention would have done if it
had been subjected to an experiment different from the one actually
performed; nothing more then this. Our quantum object possesses nothing else
than its unique and not contradictory ''instructions kit'': no two
incompatible behaviours can be adopted in the same time. The Heisenberg
principle is save.

This way of reading things dissolves the EPR paradox and has no particular
subtlety in it; furthermore it is neither particularly exotic nor confined
to the quantum domain.

To express ideas of the same kind as those presented in this paper Accardi
uses a [simpatica] metaphor: that of the chameleons \cite{camaleo}. A
chameleon is a well defined non quantum ''object'' which has the property of
assuming the colour of the support on which it rests. Now suppose you have a
pair of twin chameleons and separate them far away one from the other, then
put one of them on a leaf: it will become green. We can conclude, without
experiment, that the second chameleon, if it had been put on a leaf, \textit{%
it would have become green }too, not certainly that it \textit{is }green.
Actually suppose somebody puts the second chameleon on a trunk: it will
become brown. Nobody, knowing the result of both experiments, would conclude
that the second chameleon is green and brown at the same time. The point is
that the chameleon does not possess a colour, but the capacity to assume a
colour according to the environment in which it is laid. Everybody will
agree: 1) that the chameleons are perfectly real and existing at any moment;
2) that the phenomenon we verified is purely local; 3) that there is no
logical contradiction to be accounted for.

There is no reason that what is true for chameleons (or other classical and
macroscopic systems) cannot be true for electrons or photons. On the
contrary there are good reasons it to be true for them also.

\section{Bell inequality}

The Bell inequality has been considered as being a cornerstone in the
understanding of the nature and properties of quantum objects and phenomena.
This concept has been rather emphatically expressed in the statement by
Stapp that Bell's is ''the most profound discovery of science'' \cite{stapp}.

Bell originally applied his analysis to the EPR experiment. Maybe the
clearest way to expound it is to consider the Aspect's version of that
experiment. The core of the argument moves from the results of three sets of
measurements, two being actually performed and the third being inferred on
the basis of the original preparation of the pair of objects being tested.
The procedure is rather idealized and the physical quantities and features
of the experiment are so devised that the issues of each of the three actual
or hypothetical measurements (let us call them $a,$ $b$ and $c$) may only be 
$\pm 1$. Everything then rests on the identity \cite{peres}: 
\begin{equation}
a\left( b-c\right) \equiv \pm \left( 1-bc\right)  \label{identita}
\end{equation}

Repeating the experiment many times, then considering statistical averages
of the results one arrives at 
\begin{equation}
\left| \left\langle ab\right\rangle -\left\langle ac\right\rangle \right|
\leq 1-\left\langle bc\right\rangle  \label{disugua}
\end{equation}

which is the Bell inequality.

Using quantum mechanics to evaluate the averages one sees both by
calculation (as Bell himself did) and by experiment (Aspect and coworkers)
that in some situations (\ref{disugua}) is violated.

Now, according to Bell and to many people after him, the only hypotheses
underlying identity (\ref{identita}) are:

\begin{eqnarray}
&&a)\text{ realism}  \nonumber \\
&&b)\text{ determinism}  \label{ipote} \\
&&c)\text{ locality}  \nonumber
\end{eqnarray}
Since (\ref{identita}) proves to be false and (\ref{disugua}) violated, at
least one of the hypotheses, it is said, must be false too. The one usually
doomed to be abandoned is the third, i.e. locality (often called Einstein
locality). This introduces action at a distance and consequently violates
the principle of relativity forcing people to formulate various kinds of
arguments intended to demonstrate that the action at a distance is actually
there but cannot really be used to transmit informations at a speed higher
than that of light.

Actually however there is a fourth hidden (to say better: not declared)
hypothesis underlying Bell's reasoning: it is the assumption that the
property being measured is something possessed as such by the quantum object
and being there at any moment. This means that, even if, say, $a$ and $c$,
on one side, $b$ and $c$, on the other, cannot be measured simultaneously,
provided a rule exists enabling us to guess what the result had been if $c$
had been measured \textit{instead }of $a$ (or $c$ \textit{instead }of $b$),
we are entitled to use both $a$ and $c$ (or $b$ and $c$) at the same time in
(\ref{identita}).

The situation is different if we think quantum objects as carrying, so to
speak, recipe books rather than cakes, as said in the previous section. A
recipe produces a cake only when the right kitchen and ingredients are met
and it has no meaning to correlate a cake (= result of a measurement) with a
recipe different from the one being used. Out of metaphor, identity (\ref
{identita}) is now deprived of sense and so is the ensuing inequality (\ref
{disugua}). The logic of recipe books is different from the one of cakes.

There is nothing strange in this, nor peculiar to the quantum world, as
Accardi's chameleons show, and all three conditions (\ref{ipote}) are fully
satisfied.

\section{Conclusion}

Since the very beginning quantum mechanics subverted deeply rooted ways of
thinking and mental [abitudini]. This involved its phenomena and theory in a
halo of mystery, giving rise to a decades long debate and a host of
different interpretations of quantum phenomena calling into play free will,
the consciousness of the experimenter and things like that. Now that the
shock of the subversion of the bald programs of the XIX century physics is
far behind us, a quiet reconsidering of some of the ''strange'' behaviours
of quantum objects shows that they are neither really that strange nor
confined to the quantum world. An example is the famous EPR paradox which is
indeed no paradox, provided we accept that the state of a quantum objects is
defined by what I called an ''instruction kit'' or a ''recipe book''. In
other words, the spin or the polarization are a behaviour while interacting
rather than a colour or a label attached to a particle.

Looking at things from this viewpoint, we see that realism, determinism and
localism are generally satisfied conditions (as we reasonably expect) and
that some features of the quantum phenomena may again find their counterpart
in the macroscopic domain.

This does not mean that everything is clear and solved with quantum
mechanics: the meaning of the wave function, its collapse, the quantization
of the objects it describes still escape our imagination and, at least in my
case, our understanding. Nonetheless we may claim that part of the mystery
is only apparent and comes from a sort of long lasting prejudice.

In conclusion quantum measurements and the story of the violation of the
Bell or similar inequalities tell us that the objects of the quantum world
are not like boxes containing spin, polarization vector etc. like buttons,
pins, pearls and the like, but like programmed machines capable of different
behaviours according to the physical conditions locally triggering them. To
mock the beautiful Accardi's metaphor I would say that the quantum particles
are ''camleons'' rather than ''boxons''.

\end{document}